\documentclass[10pt]{iopart}

\usepackage{xcolor}
\usepackage{bm}
\usepackage{amssymb}
\usepackage{cite}

\definecolor{darkblue}{rgb}{0,0,0.6}
\definecolor{darkred}{rgb}{0.6,0,0}
\usepackage[colorlinks=true,urlcolor=darkblue,citecolor=darkblue, linkcolor=darkred,hyperfootnotes=false]{hyperref}

\usepackage[textsize=footnotesize]{todonotes}

\newcommand{\ind}[1]{_\mathrm{#1}}

\newcommand{\dd}{\mathrm{d}}
\newcommand{\ed}{\mathrm{e}}

\newcommand{\id}{\mathrm{i}}

\newcommand{\kk}{\boldsymbol{k}}

\newcommand{\rr}{\boldsymbol{r}}

\newcommand{\xx}{\boldsymbol{x}}

\begin{document}

\title[]{Pinning by rare defects and effective mobility for elastic interfaces in high dimensions}

\author{Xiangyu Cao$^{1}$, Vincent D\'emery$^{2,3}$, Alberto Rosso$^{4,5}$}

\address{$^1$  Department of Physics, University of California, Berkeley, Berkeley CA 94720, USA} 
\address{$^2$ Gulliver, CNRS, ESPCI Paris, PSL Research University, 10 rue Vauquelin, 75005 Paris, France}
\address{$^3$ Univ Lyon, ENS de Lyon, Univ Claude Bernard Lyon 1, CNRS, Laboratoire de Physique,
F-69342 Lyon, France}
\address{$^4$ LPTMS, CNRS, Univ. Paris-Sud, Université Paris-Saclay, 91405 Orsay, France}
\address{$^5$  Kavli Institute for Theoretical Physics, University of California, Santa Barbara, USA}

\ead{xiangyu.cao@berkeley.edu, vincent.demery@espci.fr, alberto.rosso74@gmail.com}


\begin{abstract}
The existence of a depinning transition for a high dimensional interface in a weakly disordered medium is controversial.
Following Larkin arguments and a perturbative expansion, one expects a linear response with a renormalized mobility $\mu\ind{eff}$.
In this paper, we compare these predictions with the exact solution of a fully connected model, which displays a finite critical force $f\ind{c}$.
At small disorder, we unveil an intermediary linear regime for $f\ind{c}<f<1$ characterized by the renormalized mobility $\mu\ind{eff}$.
Our results suggest that in high dimension the critical force is always finite and determined by the effect of rare impurities that is missed by the perturbative expansion.
However, the perturbative expansion correctly describes an intermediate regime that should be visible at small disorder.
\end{abstract}

%
%
%
%
%

\section{Introduction}\label{}

A  $d$-dimensional elastic interface embedded in a $d+1$ disordered medium and pulled by an external force $f$ is a paradigmatic model for the dynamics of many real systems ranging from magnetic or ferroelectric domain walls~\cite{Lemerle1998, zapperi1998dynamics, Yang1999, DurinarXiv2016} to crack ou wetting front \cite{gao1989first,maaloy2006local,BSP08,joanny1984model,rolley09}.
In these systems, elasticity and disorder compete: on one hand elastic interactions try to keep the interface flat, while impurities distort the interface.
The elastic interactions can be short ranged or decay algebraically as $1/r^{d+\alpha}$; wetting or crack fronts have $d=1$ and $\alpha=1$, domain walls have a short ranged elasticity, corresponding to $\alpha=2$, and fully connected models correspond to $\alpha=0$.

\begin{figure}
\begin{center}
\includegraphics[]{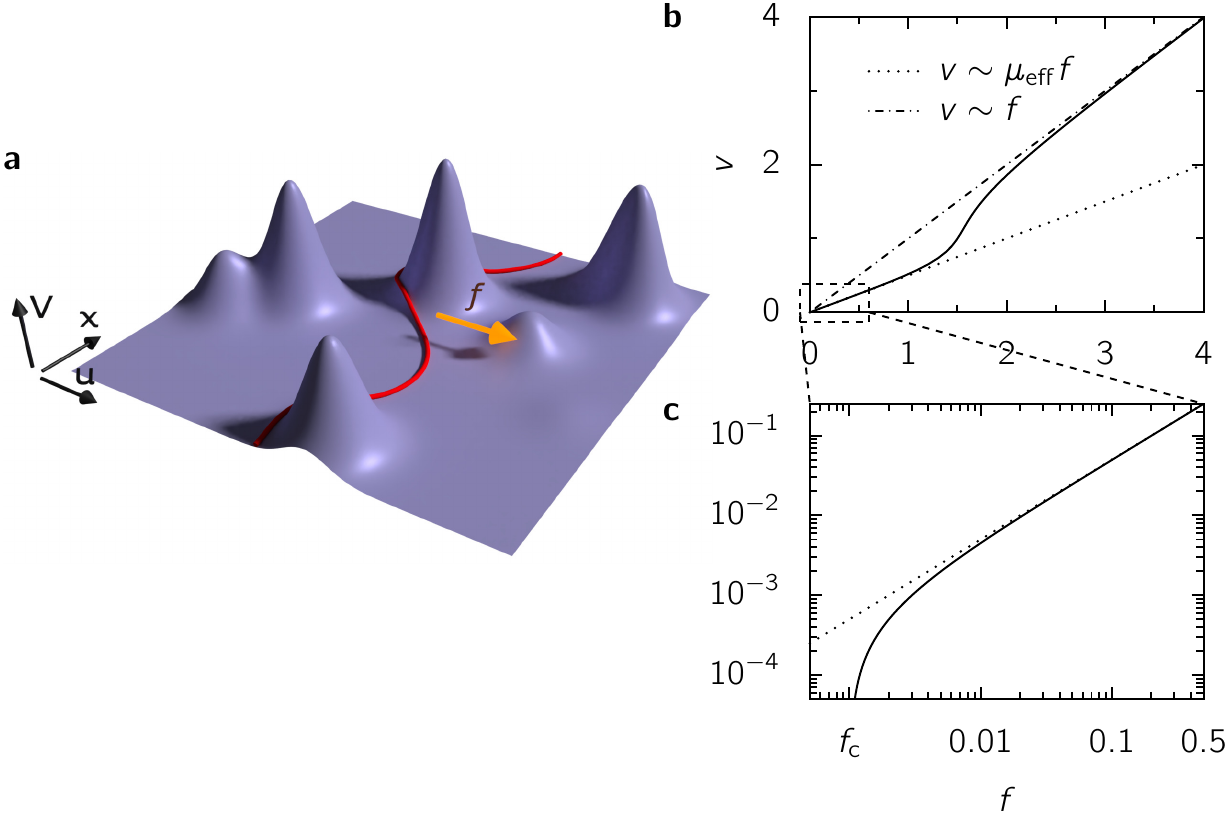}
\end{center}
\caption{{\bf a} Illustration of an elastic line in a disordered potential with strong rare defects. The disordered potential $V(\xx,u)$ depicted here generates the stochastic force $F(\xx,u) = -\partial_u V(\xx, u)$ in the equation of motion (\ref{eq:eom}).
{\bf b} Schematic velocity-force characteristics of the elastic line.
It displays a high velocity regime $v = f$ when $f \gg 1$. For $f \lesssim 1$ and at small disorder, one enters an intermediate regime $v =\mu\ind{eff} f$ with a reduced effective mobility.
{\bf c} At very small drive, $v$ departs from the linear regime and vanishes at the critical drive $f\ind{c} \ll 1$.}
\label{fig:schematic}
\end{figure}

At zero temperature, two different scenarios are expected (see figure \ref{fig:schematic}): (i) a {\em strong pinning} scenario where the interface is rough and pinned below a finite critical force, $f\ind{c}$, above which a depinning transition occurs with avalanche dynamics and non linear velocity  and (ii) a {\em weak pinning} scenario where distortions remain finite, avalanches are absent and the main effect of the disorder is to renormalize the mobility coefficient $\mu=v/f$, where $v$ is the average velocity of the interface, at small forces. 

A dimensional analysis due to Larkin~\cite{Larkin1979}  shows that,  when $d < 2 \alpha$, the interface has a finite critical force that scales as $f\ind{c}\sim \sigma^{2\alpha/(2\alpha-d)}$ for a small disorder strength $\sigma$, a prediction well confirmed numerically and analytically~\cite{Demery2014, Demery2014b, Fyodorov2017}.  
Moreover interfaces pinned by disorder should have localized soft modes, precursors of avalanche instabilities~\cite{Tanguy1998, Cao2017}. 


For $d>2\alpha$, the same argument predicts $f\ind{c}=0$. However, while there is a general consensus on the predictions in low dimension, the absence of a depinning transition for high dimensional (long range) interface is still controversial. 
The related issue of soft mode localization is also being debated~\cite{rodriguez03anderson, Cao2017}. 
For instance, the exact solution of a fully connected model with a periodic disorder, $\alpha/d=0$, shows the existence of a weak pinning phase only for a bounded disorder, while the critical force is always finite for a Gaussian disorder, in contrast with Larkin arguments~\cite{Fisher1985, Marchetti2003, Marchetti2006}.

In this paper, we study the full velocity-force characteristic of the fully-connected model and reconcile the two scenarios. The strong pinning scenario occurs at very small drive, of the order of the critical force, $f\ind{c}\sim \ed^{-1/\sigma^2}$; at stronger drive, the velocity-force characteristic is linear with an effective mobility $\mu\ind{eff}<1$ that is captured by a perturbative calculation (figure~\ref{fig:schematic}).
Note that in this regime, avalanches are not expected.

\section{Perturbative expansion}\label{sec:pert_exp}

The equation of motion of the line reads
\begin{equation}\label{eq:eom}
\partial_t u(\xx,t) = f+f\ind{el}[u(\cdot,t)](\xx)+\sigma F(\xx, u(\xx,t)),
\end{equation}
where $f$ is the external force, $F(\xx,u)$ is the stochastic force of the disorder, which is assumed to be Gaussian with correlations,
\begin{equation}
\overline{F(\rr) F(\rr')} = \Delta(\rr-\rr'),
\end{equation}
where $\rr=(\xx,u)$ and $\sigma$ is the disorder strength.
Finally, $f\ind{el}$ is the elastic force, that reads in Fourier space
\begin{equation}
\tilde f\ind{el}[u(\cdot,t)](\kk_x) = -\omega(\kk_x) \tilde u(\kk_x),
\end{equation}
with $\omega(\kk_x)=k_x^\alpha$.

The external force $f$ needed to pull the line at average velocity $v$ can be computed by placing the line in a parabolic potential of curvature $\kappa$ moving at velocity $v$, and then taking the limit $\kappa\to 0$~\cite{Demery2014b} (\ref{app:pert_calculation}).
At the lowest order in $\sigma$, we obtain:
\begin{equation}\label{eq:f_1loop}
f \simeq v + \sigma^2 \int \id k_u\tilde \Delta(\kk)G(\kk)\frac{\dd \kk}{(2\pi)^{d+1}},
\end{equation}
with the propagator
\begin{equation}\label{eq:propagator}
G(\kk)=\frac{1}{\id vk_u+\omega(\kk_x)}. 
\end{equation}
This result corresponds to the perturbative calculation at one loop.

At small velocity, the force behaves as $f\sim \mu\ind{eff}^{-1}v$, where the effective mobility $\mu\ind{eff}$ is formally given by
\begin{equation}\label{eq:mob_1loop}
\mu\ind{eff}= \left(1+\sigma^2\int \frac{k_u^2\tilde \Delta(\kk)}{|k_x|^{2\alpha}} \frac{\dd \kk}{(2\pi)^{d+1}} \right)^{-1}.
\end{equation}
The integral over $\kk_x$ converges at small $k_x$ if $d>2\alpha$. 
Note that at large $k_x$, the integral is regularized a microscopic distance such as the size of the impurity or the lattice spacing.
On the contrary, if $d\leq 2\alpha$, the integral diverges and $\mu\ind{eff}=0$, suggesting the existence of a finite critical force, which has been estimated for $d=1$ and $\alpha=1$~\cite{Demery2014b}.

The one loop result (\ref{eq:mob_1loop}) does not depend on the distribution of the disorder and is also valid for non-Gaussian disorder.
If the disorder is Gaussian, higher orders can be obtained using a diagrammatic expansion~\cite{Larkin1974} (\ref{app:pert_calculation}).
For $d>2\alpha$, they have the same scaling as the one loop result and provide further corrections to the effective mobility.

This perturbative calculation suggests that the disorder only renormalizes the mobility for $d>2\alpha$.
However, it is known that there is a finite critical force in the fully connected model ($\alpha=0$) with periodic Gaussian disorder~\cite{Fisher1985}.
Moreover, it can be argued that rare impurities result in a finite critical force for any positive values of $\alpha$ and $d$~\cite{Cao2017}.
One thus may ask if the finite value of $\mu\ind{eff}$ has a physical meaning and is observable.
In the following section, we address this question for the mean-field model with periodic Gaussian disorder using the numerical solution of the velocity-force characteristics.


\section{Mean-field model}\label{sec:mean_field}

We consider a fully connected model with periodic disorder~\cite{Fisher1985}.
The interface is discretized, and its position is given by $u_j(t)$, $j\in\mathbb{Z}$; its equation of motion is a particular form of equation~(\ref{eq:eom}), corresponding to the case $\alpha=0$:
\begin{equation}\label{eq:eom_mean_field}
\frac{\dd u_j}{\dd t}(t) = f+\langle u(t) \rangle -u_j(t) + \sigma F_j(u_j(t)),
\end{equation}
where $\langle u(t) \rangle$ is the average of position of the interface over $j$ and $\langle u(t) \rangle - u_j(t)$ plays the role of the elastic force in equation~(\ref{eq:eom}).
The disorder force is defined by
\begin{equation}
F_j(u) = -h_j\sin(u-\beta_j),
\end{equation}
where $\beta_j$ is uniformly distributed over $[0,2\pi]$ and $h_j>0$ is drawn independently from the distribution $p_h(h)=h\ed^{-h^2/2}$. 
This corresponds to a Gaussian disorder with correlation
\begin{equation}\label{eq:dis_correl_mf}
\langle F_j(u)F_{j'}(u') \rangle = \delta_{j,j'}\cos(u-u').
\end{equation}

In order to find the velocity-force characteristics, we observe that in the thermodynamic limit $\langle u(t) \rangle=vt$, where $v$ is the interface velocity.
So we  set the velocity $v$ and compute the corresponding force $f$.
For this, we first introduce a natural shift for $u_j(t)$ by defining $y_j(s)=u_j(s-[f-\beta_j]/v)-\beta_j$ which satisfies 
\begin{equation}\label{eq:eom_shifted}
\frac{\dd y^h}{\dd s}(s)=vs-y^h(s)-\sigma h\sin\left(y^h(s)\right)
\end{equation}
for $h=h_j$.
Since $y^h(s)$ evolves in a periodic potential with period $2\pi$, it should satisfy
\begin{equation}\label{eq:period_cond}
y^h\left(s+\frac{2\pi}{v}\right)=y^h(s)+2\pi.
\end{equation}
Equations (\ref{eq:eom_shifted}) and (\ref{eq:period_cond}) concern a single particle, and we solve them numerically (\ref{app:num_sol_mean_field}).
Then, we compute the force $f$ by averaging the equation of motion (\ref{eq:eom_mean_field}) over $j$, finally leading to (\ref{app:force_mf}):
\begin{equation}\label{eq:force_mf}
f = \pi- \frac{1}{T} \int_0^\infty  \int_0^T y^h(s)\dd s\, p_h(h) \dd h.
\end{equation}

\begin{figure}
\begin{center}
\includegraphics[]{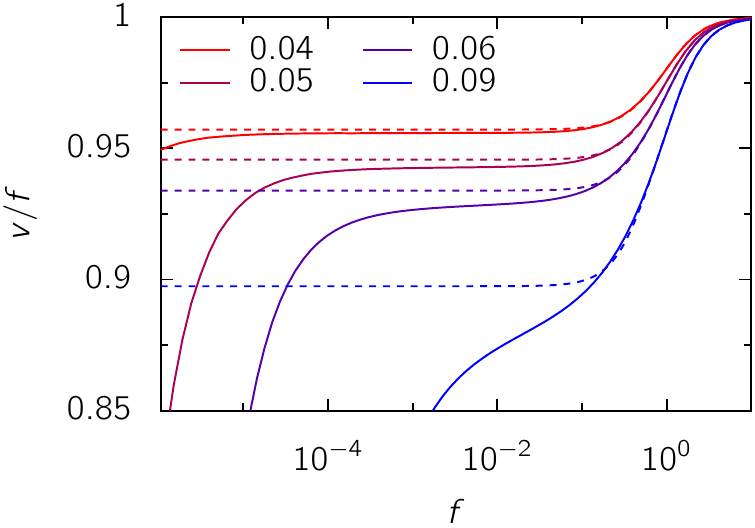}
\end{center}
\caption{Fully connected model, numerical result (solid lines) vs. two loops calculation (dashed lines, equation (\ref{eq:f4_mf})) for different disorder amplitudes $\sigma^2$.}
\label{fig:fully_connected}
\end{figure}

We compare the numerical solution with the explicit result obtained from the perturbative expansion discussed in section~\ref{sec:pert_exp}, which we apply to the mean-field model (\ref{app:mean_field_pert}):
\begin{equation}\label{eq:f4_mf}
f=v+\sigma^2\frac{v}{1+v^2}+\sigma^4\frac{v(3-7v^2-4v^4)}{(1+v^2)^3(1+4v^2)}+\mathcal{O}(\sigma^6).
\end{equation}
The numerical solution and the perturbative expansion are compared in figure~\ref{fig:fully_connected} for different disorder strength. 
Note that we plot $v/f$ as a function of $f$.

At large force, all the curves converge to a plateau which corresponds to $v=f$.
At $f\sim 1$, the perturbative calculation well describes the departure from the flow regime $v=f$.
On the contrary, the exact calculation shows that the velocity vanishes at a finite critical force, while the perturbative expansion predicts a linear response $v/f\sim \mu\ind{eff}$, with
\begin{equation}\label{eq:pert_mobility}
\mu\ind{eff}=\left(1+\sigma^2+3\sigma^4+\cdots\right)^{-1}.
\end{equation}
Interestingly, we see that at weak disorder the velocity-force characteristics develops a linear regime for $f\ind{c}\lesssim f\lesssim 1$.
This regime is characterized by an effective mobility, which is given by the height of the plateau in figure~\ref{fig:fully_connected} and is captured by the perturbative expansion (equation~(\ref{eq:pert_mobility})).

To further understand the results, let us remind that the critical force can be computed exactly for the mean-field model~\cite{Fisher1985} and reads
\begin{equation}\label{eq:fc_mf}
f\ind{c}=\int_{1/\sigma}^\infty f\ind{c}^{\sigma h} p_h(h)\dd h,
\end{equation}
where $f\ind{c}^{\sigma h}$ is the critical force in the mean-field model where $h$ is fixed and $\beta$ is uniformly distributed over $[0,2\pi]$. Exact calculations~\cite{Fisher1985,Cao2017} show that $f\ind{c}^{\sigma h}=0$ for $\sigma h<1$, while $f\ind{c}^{\sigma h}>0$ for $\sigma h>1$, with the asymptotic behavior $f\ind{c}^{\sigma h}\sim (\sigma h-1)^2$ near $\sigma h\to 1^+$. Thus, in the Gaussian model with small $\sigma$, the critical force (\ref{eq:fc_mf}) is dominated by the rare impurities with $\sigma h\simeq 1^{+}$, and
\begin{equation}
f\ind{c}\sim_{\sigma\to 0} \sigma^4\ed^{-\frac{1}{2\sigma^2}}.
\end{equation}
Since the Taylor series of this expression around $\sigma=0$ is zero, the perturbative expansion cannot capture the effect of these rare impurities and does not predict the existence of the finite critical force.

\section{Conclusion}\label{}

In this article, we have computed the velocity-force characteristics of a fully connected model ($\alpha=0$) with Gaussian disorder using two approaches: the exact solution, and a perturbative expansion at small disorder.
We have shown that the exact solution has three main features: (i) a finite critical force $f\ind{c}$, (ii) an intermediate linear regime with an effective mobility $\mu\ind{eff}<1$, and (iii) a high velocity regime where $v=f$.
While the perturbative expansion misses the existence of the critical force, it correctly describes the features (ii) and (iii); in particular, it provides a good estimation of $\mu\ind{eff}$ at small disorder.

For $\alpha>0$, there is no exact solution for the velocity-force characteristics.
However, we dispose of the perturbative expansion and of numerical simulations.
Our results suggest the following scenario:
\begin{itemize}
\item If $d>2\alpha$, the perturbative expansion predicts a linear response at small $\sigma$. 
We expect that the model presents a finite critical force generated by rare impurities. The value of the critical force is not analytical in $\sigma$ as $\sigma\to 0$ and depends on the disorder distribution.
On the contrary, the effective mobility predicted by the perturbative expansion should be visible when $f\ind{c}\lesssim f\lesssim 1$ and does not depend on the disorder distribution at order $\sigma^2$.
\item If $d<2\alpha$, the critical force is always finite and at small disorder it is determined by typical impurities, and the Larkin arguments predict $f\ind{c}\sim \sigma^{\frac{2\alpha}{2\alpha -d}}$.
The proportionality constant of the critical force could be evaluated analytically, either through the perturbative expansion presented here, which has been done for $\alpha=1$, $d=1$~\cite{Demery2014b}, or through a Kac-Rica approach, which has been done for $\alpha=2$ and $d=1$~\cite{Fyodorov2017}.
\end{itemize}

\textit{Acknowledgments.} The authors acknowledge support from a Simons Investigatorship, Capital Fund Management Paris and LPTMS (X.C.), and the ANR grant ANR-16-CE30-0023-01 THERMOLOC (A.R.).  This research was supported in part by the National Science Foundation under Grant No. NSF PHY 17-48958.

\appendix

\section{Perturbative calculation of the force}\label{app:pert_calculation}

In order to find the force required to pull the interface at an average velocity $v$, we replace the force in (\ref{eq:eom}) by a moving parabolic well:
\begin{equation}\label{eq:eom_parabola}
\partial_t u(\xx,t) = \kappa[vt-u(\xx,t)]+f\ind{el}[u(\cdot,t)](\xx)+\sigma F(\xx,u(\xx,t)).
\end{equation}
We obtain the force as the average force exerted by the parabola on the interface:
\begin{equation}
f = \kappa\left[vt - \overline{u(\xx,t)}\right].
\end{equation}

We want to solve perturbatively the equation~(\ref{eq:eom_parabola}). Thus we expand $u(\xx,t)$ as follows:
\begin{equation}
u(\xx,t)=\sum_{n=0}^\infty \sigma^n u_n(\xx,t).
\end{equation}
Plugging this expansion into the Taylor expansion of the random force $F(\xx,u)$ around $u_0(t)=v(t-\kappa^{-1})$ leads to
\begin{eqnarray}
F(\xx,u(\xx,t)) &= \sum_{k=0}^\infty \frac{1}{k!} \left(\sum_{n=1}^\infty \sigma^n u_n(\xx,t) \right)^k\partial_u^k F(\xx,u_0(t))\\
& = \sum_{n=0}^\infty \sigma^n \sum_{\lambda\in \mathcal{P}_n}\frac{\partial_u^{|\lambda|} F(\xx,u_0(t))}{s_\lambda}\prod_{i=1}^{|\lambda|} u_{\lambda_i}(\xx,t),
\end{eqnarray}
where $\mathcal{P}_n$ is the set of partitions of the integer $n$, $\lambda=(\lambda_i)\in\mathcal{P}_n$ is a partition of $n$ with $\lambda_1\geq\dots\geq\lambda_{|\lambda|}$, $|\lambda|$ is the number of elements in this partition, and $s_\lambda$ is the number of permutations that leave the partition invariant.

The equation of evolution of the order $n\geq 1$ is thus given by
\begin{equation}\label{eq:edp_un}
\fl\partial_t u_n(\xx,t)=-\kappa u_n(\xx,t)+f\ind{el}[u_n(\cdot,t)](\xx)
+\sum_{\lambda\in \mathcal{P}_{n-1}}\frac{\partial_u^{|\lambda|} F(\xx,u_0(t))}{s_\lambda}\prod_{i=1}^{|\lambda|} u_{\lambda_i}(\xx,t).
\end{equation}
In order to integrate easily the linear terms (all the terms except the third on the r.h.s), we write the order $n$ as
\begin{equation}
u_n(\xx,t)= \int\tilde u_n(\kk)\ed^{\id[\kk_x\cdot\xx + k_u u_0(t)]}\frac{\dd\kk}{(2\pi)^{d+1}},
\end{equation}
where we recall that $\kk=(\kk_x, k_u)$.
Note that the integral over $k_u$ is equivalent to a Fourier transform in time; we choose this definition because it simplifies more easily.
Inserting this expression in equation (\ref{eq:edp_un}), we get
\begin{equation}\label{eq:un_convol}
\tilde u_n(\kk)=G'(\kk) \sum_{\lambda\in \mathcal{P}_{n-1}} \frac1{s_\lambda} \left(\left[(\id k_u)^{|\lambda|}\tilde F \right]*\tilde u_{\lambda_1} *\cdots *\tilde u_{\lambda_{|\lambda|}}\right)(\kk),
\end{equation}
where $G'(\kk)=[\id vk_u+\kappa+\omega(\kk_x)]^{-1}$ and each convolution (denoted by $*$) involves an integration over a moment with a factor $(2\pi)^{-d-1}$.
This expression is easier to visualize using a diagrammatic representation.

The sum over the partitions of $n-1$ in equation (\ref{eq:un_convol}) can be seen as a sum of diagrams. 
In the diagram associated to the partition $\lambda$, there are $|\lambda|$ lower order terms $\tilde u_{\lambda_i}$ entering a vertex with moments $\kk_i$ and a disorder term (the term in brackets) entering the vertex with moment $\kk_0$.
A propagator with weight $G'(\kk)$ goes out of the vertex, and the momentum is conserved at the vertex: $\kk=\sum_{i=0}^{|\lambda|}\kk_i$.
Finally, the symmetry factor $s_\lambda$ is the number of permutations of the entering $\tilde u_{\lambda_i}$ leaving the diagram invariant.

\begin{figure}
\begin{center}
\includegraphics[scale=.25]{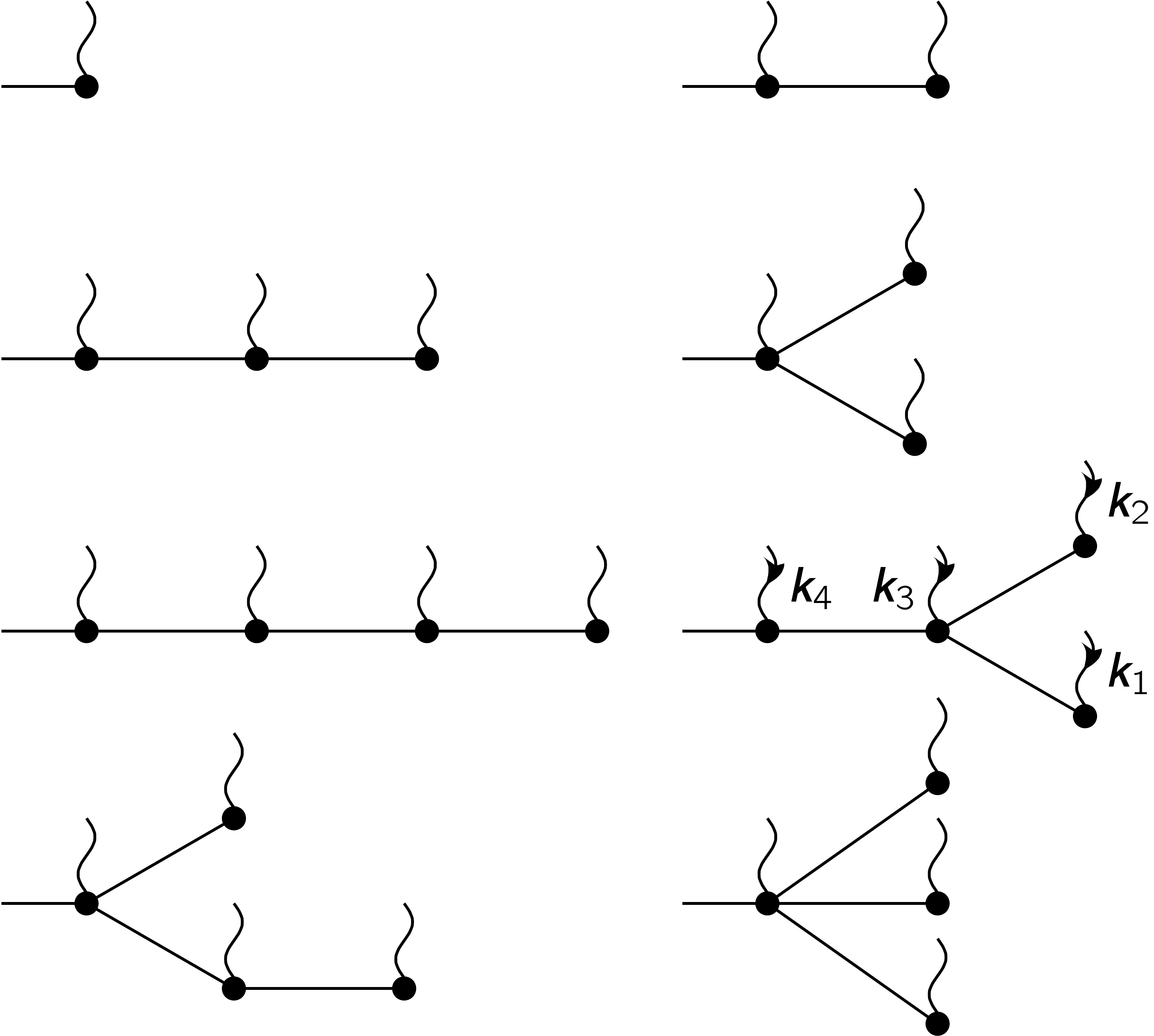}
\end{center}
\caption{Backbones for the orders 1 and 2 (first line), 3 (second line), and 4 (third and fourth lines).
Their symmetry factors are, from left to right and top to bottom: 1, 1, 1, $1/2$, 1, $1/2$, 1, $1/6$.
The momenta carried by the incoming disorder lines are labeled for the diagram evaluated in equation (\ref{eq:contrib_backbone}).}
\label{fig:backbones}
\end{figure}

The order $n$ involves all the orders $k<n$.
In turn, the order $k<n$ can be expressed with orders $l<k$.
By recurrence, the order $n$ is given as a function of the order $0$; and is conveniently computed using the diagrams presented above.
First, the diagrams at order $n$ are drawn according to the following rules (figure~\ref{fig:backbones}):
\begin{itemize}
\item There are $n$ vertices, with a single leftmost vertex.
\item Each vertex has a single straight line (propagator) flowing out to the left, any number of propagators flowing in from the right, and a single wavy line (disorder line) flowing in.
\item Each propagator has to emerge out of a vertex.
\end{itemize}
The diagrams, obtained before the average over disorder, are called the backbones; there is $1$ backbone at order $1$, $1$ backbone at order $2$, $2$ backbones at order $3$, and $4$ backbones at order $4$.
The value associated to a diagram for a momentum $\kk$ flowing out is calculated with
\begin{itemize}
\item Each line carries a momentum $\kk$, the propagators come with a weight $G'(\kk)$ and the disorder lines with a weight $(\id k_u)^m\tilde F(\kk)$, where $m$ is the number of propagators flowing in the vertex.
\item The momenta entering the diagram through the disorder lines are integrated over with a weight $(2\pi)^{-d-1}$.
\item The momentum conservation is enforced at each vertex using a delta function $(2\pi)^{d+1}\delta(\kk\ind{out}-\sum_i \kk_{\mathrm{in},i}) $, where $\kk\ind{out}$ is the momentum flowing out of the vertex and $\kk_{\mathrm{in},i}$ are the momenta flowing in the vertex, either through propagators or through a disorder line.
\item There is a symmetry factor given by the inverse of the number of permutations of the propagators entering each vertex that leave the diagram invariant.
\end{itemize}
For instance, the second diagram on the third line in figure~\ref{fig:backbones}, has the value
\begin{equation}\label{eq:contrib_backbone}
\fl\frac{G'(\kk)}{2}\int (\id k_{u3})^2(\id k_{u4})G'(\kk_1)G'(\kk_2)G'(\kk_1+\kk_2+\kk_3)\delta\left(\kk-\sum_{i=1}^4\kk_i\right)\frac{\prod_{i=1}^4\tilde F(\kk_i)\dd\kk_i}{(2\pi)^{3(d+1)}}.
\end{equation}
The symmetry factors of the diagrams up to order 4 are given in the caption of figure~\ref{fig:backbones}.

\begin{figure}
\begin{center}
\includegraphics[scale=.25]{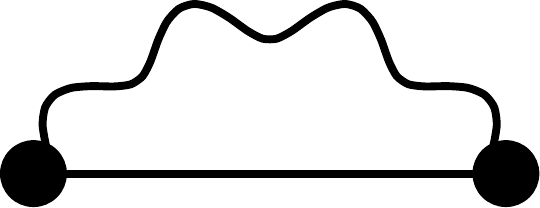}
\end{center}
\caption{One loop diagram giving the force at order $\sigma^2$.}
\label{fig:diagram_1loop}
\end{figure}

\begin{figure}
\begin{center}
\includegraphics[scale=.25]{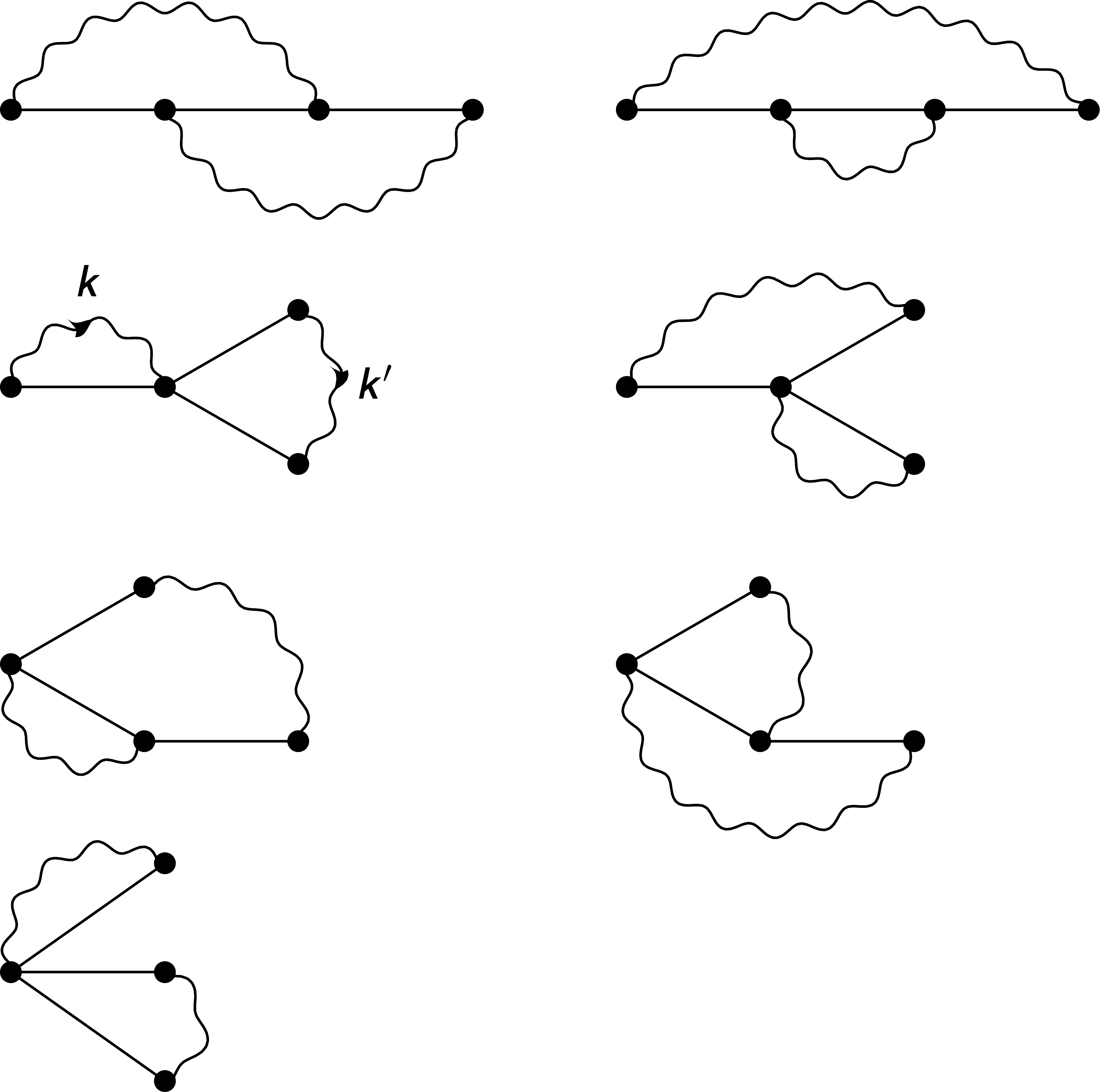}
\end{center}
\caption{Two loops diagrams giving the force at order $\sigma^4$.
The momenta carried by the disorder correlators are labeled for the diagram evaluated in equation (\ref{eq:contrib_2loops}).}
\label{fig:diagrams_2loops}
\end{figure}

The final step is to average over disorder. 
The average over disorder of a backbone is obtained by pairing the disorder lines in all the possible ways.
The terms $\tilde F(\kk)$ of the disorder lines are replaced by the disorder correlators $\tilde \Delta (\kk)$; note that the momenta entering the end vertices of a disorder line carrying a momentum $\kk$ are $\kk$ and $-\kk$.
The symmetry factor of a diagram is still given by the number of permutations of the propagators entering each vertex that leave the diagram invariant, but it is affected by the pairings.
There is a single diagram with one loop at order 2 (figure~\ref{fig:diagram_1loop}), and seven diagrams with two loops at order 4 (figure~\ref{fig:diagrams_2loops}).
Note that two ``one particle reducible'' diagrams with two loops are omitted; the reason is given below.

Due to momentum conservation, the delta function at the leftmost vertex reads $(2\pi)^{d+1}\delta(\kk)$, and it can be factorized.
This corresponds to the fact that $\overline{u_n(\xx,t)}$ does not depend on $\xx$ or $t$.
The propagator flowing out of this vertex to the left thus carries a momentum $\kk=0$, which has a contribution $G'(0)=\kappa^{-1}$; this is the reason why we do not represent this propagator on the diagrams.
The force at order $n$ is given by $f_n=-\kappa \bar u_n$, and the $\kappa$ cancels the contribution of the outgoing propagator. 
In the limit of a weak parabolic trap, $\kappa\to 0$, the internal propagators $G'(\kk)$ are replaced by $G(\kk)$.
If a diagram is one particle reducible, it has a propagator carrying a momentum $\kk=0$, with a contribution $\kappa^{-1}$ that diverges when $\kappa\to 0$.
It can be shown that the contributions of the two two-loops diagrams that are one particle reducible vanish.

To summarize, the contribution of one-loop and two-loops diagrams given in Figs.~\ref{fig:diagram_1loop} and \ref{fig:diagrams_2loops} are computed with the rules:
\begin{itemize}
\item Each line carries a momentum $\kk$, and the momentum is conserved at the vertices.
\item A straight line (propagator) with momentum $\kk$ comes with a factor $G(\kk)$, and the disorder correlator with a factor $\tilde \Delta (\kk)$.
\item Each vertex has a factor $(\id k_u)^m$, where $k_u$ is the $u$-component of the momentum coming in from the disorder correlator, and $m$ is the number of propagators entering the vertex from the right.
\item The free momenta (one per loop) are integrated over with a factor $(2\pi)^{-d-1}$.
\item There is a symmetry factor given by the number of permutations of the propagators entering each vertex that leave the diagram invariant, and a global minus sign.
\end{itemize}

For instance, the contribution of the one-loop diagram (figure~\ref{fig:diagram_1loop}) is given by
\begin{equation}
-\int (-\id k_u) G(\kk)\tilde\Delta(\kk)\frac{\dd\kk}{(2\pi)^{d+1}},
\end{equation}
which gives equation~(\ref{eq:f_1loop}).
The contribution of the first diagram of the second row in figure~\ref{fig:diagrams_2loops} is 
\begin{equation}\label{eq:contrib_2loops}
-\frac{1}{2}\int (-\id k_u)(\id k_u)^2 G(\kk)G(\kk')G(-\kk')\tilde \Delta(\kk)\tilde \Delta(\kk')\frac{\dd\kk\dd\kk'}{(2\pi)^{2(d+1)}}.
\end{equation}

\section{Numerical solution of equation (\ref{eq:eom_shifted})}\label{app:num_sol_mean_field}

\begin{figure}
\begin{center}
\includegraphics[width=.6\linewidth]{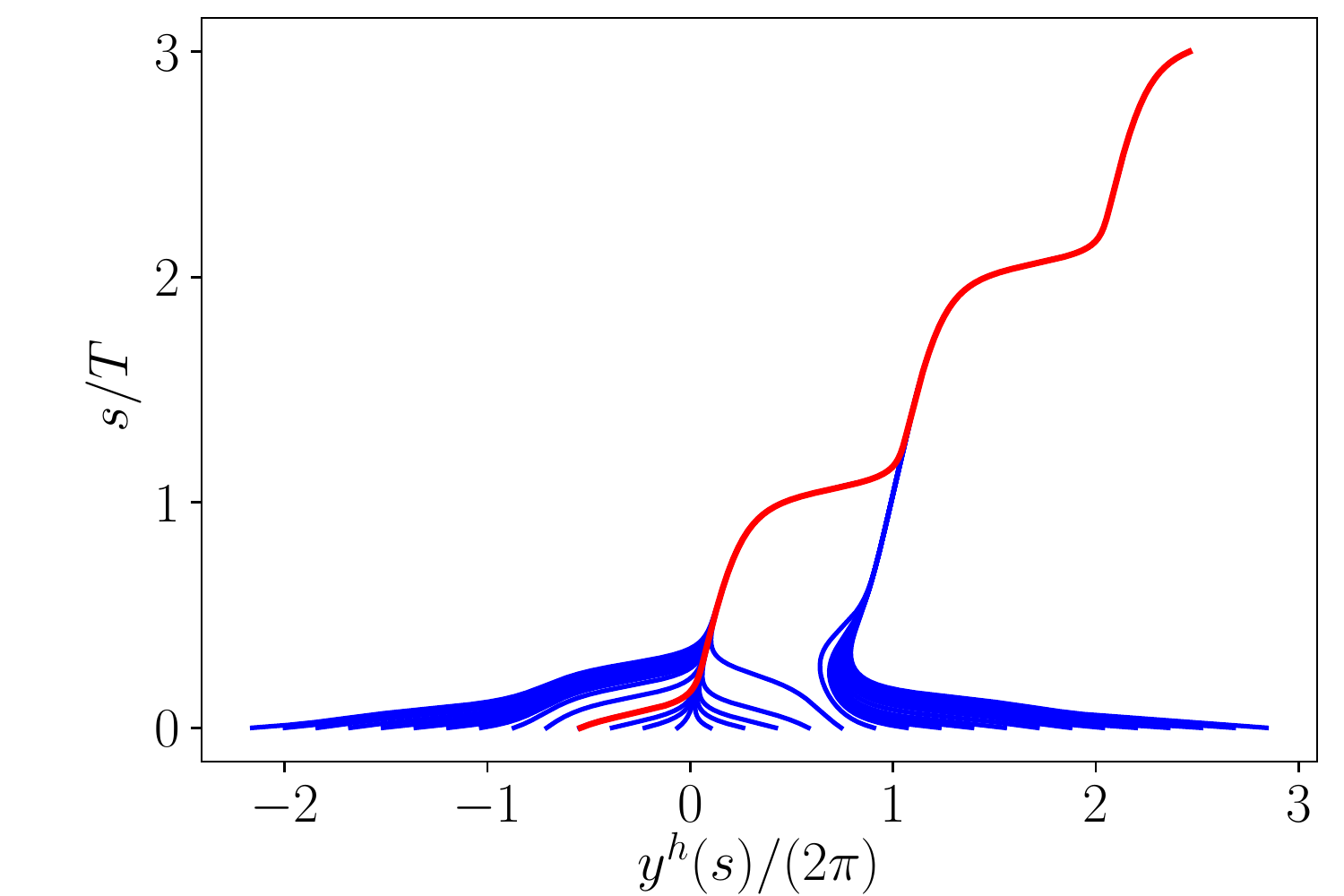}
\end{center}
\caption{Trajectories of the differential equation (\ref{eq:eom_shifted}) over three periods with $\sigma h=3$ and $v=1$. Different thin curves correspond to different initial conditions; the emanating trajectories converge to the unique solution satisfying the periodicity condition (\ref{eq:period_cond}) (thick curve).
}
\label{fig:num_sol_mf_dynamic}
\end{figure}

For a given $v$ and $\sigma h$, the numerical solution of equation (\ref{eq:eom_shifted}) is obtained by a standard fourth-order Runge-Kutta scheme. Starting from an arbitrary initial condition at time $s=0$, $y^h(s)$ quickly converges to solution satisfying the periodicity condition (\ref{eq:period_cond}), as shown in figure \ref{fig:num_sol_mf_dynamic}. 

\section{Derivation of equation (\ref{eq:force_mf})}\label{app:force_mf}

Averaging the equation of motion (\ref{eq:eom_mean_field}) over $j$ and using $\langle u_j(t) \rangle=vt$ leads to
\begin{equation}
f=v+\sigma \left\langle h_j\sin(u_j(t)-\beta_j) \right\rangle;
\end{equation}
we note that the three terms are all time independent.
Taking the average over one period $T=2\pi/v$ gives
\begin{eqnarray}
f & = v + \frac{\sigma}{T} \left\langle h_j\int_0^T \sin(u_j(t)-\beta_j)\dd t \right\rangle\\
& = v + \frac{\sigma}{T} \left\langle h_j\int_0^T \sin(y^{h_j}(s))\dd s \right\rangle,
\end{eqnarray}
where we used the periodicity condition $y_j(s+T)=y_j(s)+2\pi$.
In the thermodynamic limit, the average over $j$ corresponds to an average over $h$, leading to
\begin{equation}\label{eq:f1}
f = v + \frac{\sigma}{T} \int_0^\infty  h\int_0^T \sin\left(y^h(s)\right)\dd s\, p_h(h) \dd h.
\end{equation}
Finally, averaging the equation (\ref{eq:eom_shifted}) over a period leads to
\begin{equation}
\frac{\sigma h}{T}\int_0^T\sin \left(y^h(s) \right)\dd s = \pi - v - \frac{1}{T}\int_0^T y^h(s)\dd s.
\end{equation}
Using this relation in equation (\ref{eq:f1}), we get equation (\ref{eq:force_mf}).

\section{Application of the perturbative calculation to the mean-field model}\label{app:mean_field_pert}

Here, we apply the perturbative calculation presented in~\ref{app:pert_calculation} to the mean-field model defined in Sec.~\ref{sec:mean_field}.
First, the interface is defined on $\mathbb{Z}$, meaning that $k_x\in[-\pi,\pi]$.
The mean-field model corresponds to $\alpha=0$, hence the propagator is given by $G(\kk)=(1+\id vk_u)^{-1}$.
The disorder correlator is given by equation~(\ref{eq:dis_correl_mf}), and its Fourier transform reads $\tilde \Delta(\kk)=\pi[\delta(k_u-1)+\delta(k_u+1)]$.
Due to this particular form,  the integration over the momentum $k_u$ going through a disorder correlator is a sum over $\{-1,1\}$, with a factor $1/2$.
Since nothing depends on the $x$ component $k_x$ of the wavevector, the integration over it and the division by a factor $2\pi$ gives 1.

\begin{figure}
\begin{center}
\includegraphics[]{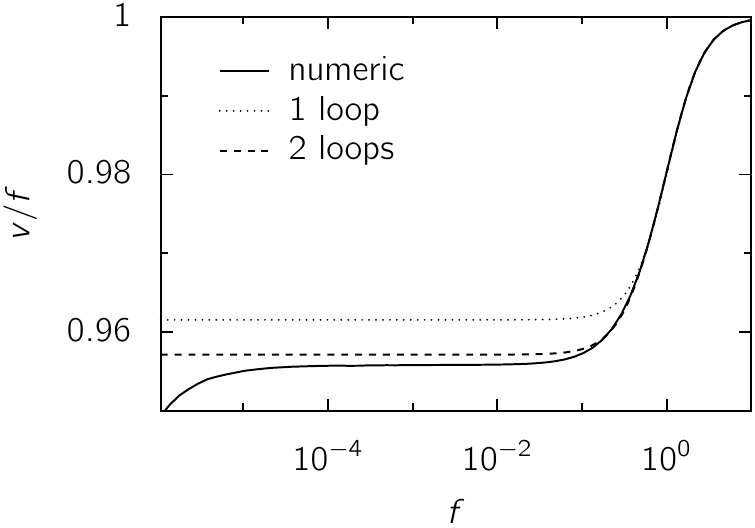}
\end{center}
\caption{One loop and two loops perturbative calculation of the velocity-force characteristics for the fully connected model at $\sigma^2=0.04$ (equations~(\ref{eq:mf_1loop}, \ref{eq:mf_2loops})).
}
\label{fig:fully_connected_loops}
\end{figure}

We can now evaluate the force velocity characteristic of the mean-field model to two loops.
The one-loop contribution to the force is:
\begin{equation}\label{eq:mf_1loop}
f_2 =  - \frac{1}{2}\sum_{k_u\in\{-1,1\}} (-\id k_u) \frac{1}{1+\id vk_u} = \frac{v}{1+v^2}.
\end{equation}
The contributions of the two-loops diagrams shown in figure~\ref{fig:diagrams_2loops} are, from left to right and top to bottom:
\begin{eqnarray}
f_4^{(1)} & = \frac{2v-v^3}{(1+v^2)^2(1+4v^2)},\\
f_4^{(2)} & = \frac{3v}{(1+v^2)(1+4v^2)},\\
f_4^{(3)} & = -\frac{v}{2(1+v^2)^2},\\
f_4^{(4)} & = -\frac{2v}{(1+v^2)^3},\\
f_4^{(5)} & = \frac{v}{(1+v^2)(1+4v^2)},\\
f_4^{(6)} & = -\frac{3v^3}{(1+v^2)^2(1+4v^2)},\\
f_4^{(7)} & = -\frac{v}{2(1+v^2)^2}.
\end{eqnarray}
Summing all these contributions finally gives
\begin{equation}\label{eq:mf_2loops}
f_4 = \frac{v(3-7v^2-4v^4)}{(1+v^2)^3(1+4v^2)} \,.
\end{equation}
Hence the force up to order $\sigma^4$ is given in equation~(\ref{eq:f4_mf}).
The perturbative expansion at order $\sigma^2$ and $\sigma^4$ is compared to the exact result in figure~\ref{fig:fully_connected_loops}.

\section*{References}

\bibliographystyle{iopart-num}

\bibliography{biblio}

\end{document}